\documentclass[aps,prl,reprint,superscriptaddress,footinbib]{revtex4-1}

\usepackage{graphicx}
\usepackage{dcolumn}
\pdfoutput=1

\begin{document}

\title{Number fluctuations of sparse quasiparticles in a superconductor} 

\author{P.J. de Visser}
\email{p.j.devisser@tudelft.nl}

\affiliation{SRON National Institute for Space Research, Sorbonnelaan 2, 3584 CA Utrecht, The Netherlands}
\affiliation{Kavli Institute of NanoScience, Faculty of Applied Sciences, Delft University of Technology, Lorentzweg 1, 2628 CJ Delft, The Netherlands}

\author{J.J.A. Baselmans}
\author{P. Diener}
\author{S.J.C. Yates}
\affiliation{SRON National Institute for Space Research, Sorbonnelaan 2, 3584 CA Utrecht, The Netherlands}

\author{A. Endo}
\author{T.M. Klapwijk}
\affiliation{Kavli Institute of NanoScience, Faculty of Applied Sciences, Delft University of Technology, Lorentzweg 1, 2628 CJ Delft, The Netherlands}

%\date{\today}

\begin{abstract}

We have directly measured quasiparticle number fluctuations in a thin film superconducting Al resonator in thermal equilibrium. The spectrum of these fluctuations provides a measure of both the density and the lifetime of the quasiparticles. We observe that the quasiparticle density decreases exponentially with decreasing temperature, as theoretically predicted, but saturates below 160 mK to 25-55 /$\mu$m$^3$. We show that this saturation is consistent with the measured saturation in the quasiparticle lifetime, which also explains similar observations in qubit decoherence times.

\end{abstract}

%\pacs{74.40.-n, 74.25.Bt, 74.25.N-, 07.57.Kp}% insert suggested PACS numbers in braces on next line

\maketitle

In a superconductor the density of unpaired electrons (quasiparticles) should vanish when approaching zero temperature \cite{thinkham}. This crucial property promises long decoherence times for superconducting qubits \cite{jclarke2008} and long relaxation times for highly sensitive radiation detectors \cite{pday2003}. However, relaxation times for resonators  \cite{rbarends2008c,rbarends2009} and qubit decoherence times \cite{jaumentado2004,mshaw2008,jmartinis2009} were shown to saturate at low temperature. Recent modeling \cite{jmartinis2009,gcatelani2011} suggests that non-equilibrium quasiparticles are the main candidate for this saturation, which was tested qualitatively by injecting quasiparticles into a qubit \cite{mlenander2011}. A direct measurement of the number of quasiparticles and the energy decay rate in equilibrium at low temperatures would provide new insight in superconductivity at low temperatures, crucially needed in the aforementioned fields. 

At finite temperature, it follows from thermodynamics that the density of quasiparticles fluctuates around an average value that increases exponentially with temperature \cite{cwilson2004}. Here we report a measurement of the spectrum of these fluctuations in a single aluminium superconducting film ($T_c$ = 1.1 K) in equilibrium, for temperatures from 300 mK to 100 mK. The number fluctuations show up as fluctuations in the complex conductivity of the film, probed with a microwave resonator. The spectrum of these fluctuations provides a direct measure of the number of quasiparticles in the superconductor. We observe that the quasiparticle density decreases exponentially with decreasing temperature until it saturates at 25-55 $\mu$m$^{-3}$ below 160 mK. We prove that the measured saturation of the quasiparticle lifetime to 2.2 ms below 160 mK is consistent with the saturation in quasiparticle density. Besides the fundamental significance, our experiment shows that it is possible to reach the fundamental generation-recombination noise limit in microwave kinetic inductance detectors based upon Al resonators.

\begin{figure}

\includegraphics{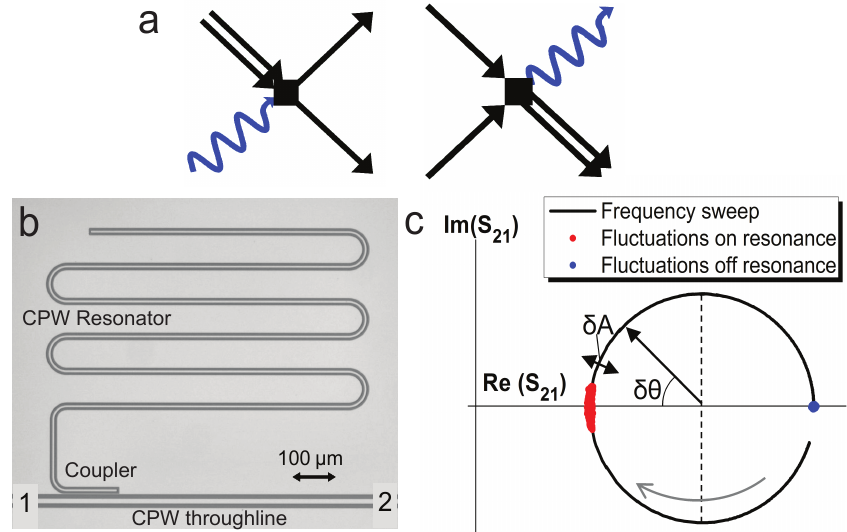} %
\caption{\label{fig:deviceandcircle} (color online) (a) Schematic of the process of generation (left) and recombination (right) of quasiparticles in a superconductor. The single arrow symbolises a quasiparticle, the double arrow a Cooper pair and the wavy arrow a phonon. (b) A microscope picture of the microwave resonator, two times reduced in length for visibility. The open ends set the half wavelength resonance condition. The coplanar waveguide (CPW) through line is used for excitation and readout (from contact 1 to 2) of the resonator with a microwave signal. (c) The real and imaginary parts of the complex transmission $S_{21}$ as a function of frequency around the resonator resonant frequency. The gray arrow indicates the direction of increasing frequency. The $S_{21}$ traces out a circle in the complex plane, off-set from the origin, around the resonant frequency of the resonator. We define a resonator amplitude $A$ and phase $\theta$ with respect to the resonance circle centre as indicated. In red, the measured fluctuations in $S_{21}$ are shown in case the readout frequency is fixed at the resonant frequency. The fluctuations far off-resonance are shown in blue, which are used as a calibration for system noise contributions.}

\end{figure}

In a superconductor in thermal equilibrium, the density of quasiparticles per unit volume is given by
\begin{equation}
	n_{qp} = 2N_0\sqrt{2\pi k_BT\Delta}\exp(-\Delta/k_BT),
	\label{eq:Nqp}
\end{equation}
valid at $k_BT < \Delta$, with $N_0$ the single spin density of states at the Fermi level ($1.72\times 10^{10}$ $\mu$m$^{-3}$eV$^{-1}$ for Al), $k_B$ Boltzmann's constant, $T$ the temperature and $\Delta$ the energy gap of the superconductor. Two quasiparticles with opposite spins and momenta can be generated from a Cooper pair by a phonon with an energy larger than the energy gap. When two quasiparticles recombine into a Cooper pair, a phonon is emitted. These processes, schematically depicted in Fig. \ref{fig:deviceandcircle}a, are random processes in equilibrium. Assuming a thermal distribution of quasiparticles and phonons at low temperature, the average quasiparticle recombination time is given by \cite{skaplan1976}
\begin{equation}
	\tau_r=\frac{\tau_0}{\sqrt{\pi}}\left(\frac{k_BT_c}{2\Delta}\right)^{5/2}\sqrt{\frac{T_c}{T}}\exp(\Delta/k_BT)=\frac{\tau_0}{n_{qp}}\frac{2N_0(k_BT_c)^3}{(2\Delta)^2},
	\label{eq:taur}
\end{equation}
where $T_c$ is the critical temperature of the superconductor and $\tau_0$ a material dependent, characteristic electron-phonon interaction time. Eqs. \ref{eq:Nqp} and \ref{eq:taur} predict a very low quasiparticle density and consequently a very long quasiparticle lifetime at temperatures $T<T_c/10$. 

The process of random generation and recombination of charge carriers is a well-studied phenomenon in solid state physics, in particular in semiconductors, but has hardly been studied in superconductors. In one earlier experiment, this generation-recombination noise was identified in the current fluctuations through a tunnel barrier connected to a small Al quasiparticle box \cite{cwilson2001}, although only down to an intermediate 210 mK. The general theory of quasiparticle number fluctuations in a superconductor \cite{cwilson2001,cwilson2004} connects the frequency dependence of the fluctuations to the microscopic dynamics of quasiparticle generation and recombination. The dominant timescale of these processes is the recombination time of a quasiparticle, ($\tau_r$ is about 1 ms in an Al film \cite{rbarends2008c}), because the phonon pair breaking time and the phonon escape time are both much shorter, about $10^{-10}$ s based on a film thickness of 40 nm \cite{skaplan1976,skaplan1979}. In thermal equilibrium, the generation and recombination rates are equal and the variance of the random number fluctuations $\sigma^2=<\delta N_{qp}^2>=N_{qp}=n_{qp}V$, with $V$ the volume of the system. The power spectral density of these fluctuations shows a Lorentzian spectrum, given by
\begin{equation}
	S_N(\omega)=\frac{4N_{qp}\tau_r}{1+(\omega\tau_r)^2},
	\label{eq:SN}
\end{equation}
with $\omega$ the angular frequency. Eqs. \ref{eq:Nqp} and \ref{eq:taur} show that the product $N_{qp}\tau_r$ is constant over temperature, whereas the total integrated power spectral density increases exponentially with temperature. This is because the bandwidth of the fluctuations increases exponentially with temperature as well, as it scales with $1/\tau_r^2$. We emphasise that this property is unique for quasiparticle generation-recombination noise in a superconductor.

We measure the quasiparticle number fluctuations using a high-quality microwave resonator. The high frequency response of the superconductor is controlled by the quasiparticle density through the complex conductivity $\sigma_1-i\sigma_2$. The real part, $\sigma_1$, is resistive and denotes the conductivity by quasiparticles. The imaginary part, $\sigma_2$, is inductive and due to the superconducting condensate, the Cooper pairs \cite{dmattis1958}. Quasiparticle number fluctuations will show up as fluctuations in the complex conductivity. To measure the complex conductivity, a 40 nm thick Al film was patterned into microwave resonators. The film was sputtered deposited onto a C-plane sapphire substrate. The critical temperature is 1.11 K, from which the energy gap $\Delta=1.76k_BT_c=168$ $\mu$eV. The low temperature resistivity $\rho=0.8$ $\mu\Omega$cm and the residual resistance ratio RRR = 5.2. The film was patterned by wet etching into distributed, half wavelength, coplanar waveguide resonators, with a defined central line width of 3.0 $\mu$m and gaps of 2.0 $\mu$m wide (Fig. \ref{fig:deviceandcircle}b). The resonator under consideration shows its lowest order resonance at 6.61924 GHz and has a central strip volume of $1.0\cdot 10^3$ $\mu$m$^3$. The low temperature (100 mK) resonance curve shows a coupling limited quality factor of 3.87$\times 10^4$. The samples are cooled in a pulse tube pre-cooled adiabatic demagnetization refrigerator. The cold stage is surrounded by a superconducting magnetic shield inside a cryoperm shield. Special care has been taken to make the setup light tight, such that no excess quasiparticles are created by stray light. The sample is mounted inside a light tight holder which itself is placed inside another light tight box (also at base temperature), to prevent radiation leaking in via the coax cable connectors. Radiation absorber, consisting of a mixture of carbon black, epoxy and SiC grains is placed inside both the sample holder and the outer box. The outer box is equipped with special coax cable filters that attenuate all frequencies above 10 GHz exponentially (see Ref. \onlinecite{jbaselmans2009} for details). The system is proven to be light tight by measuring the quasiparticle lifetime as a function of the 4 K-stage temperature while keeping the sample at base temperature. Within the measurement accuracy (15\%), there was no change in the lifetime, indicating that the stray-light power at the chip is negligible. For the lifetime measurements the sample is illuminated with a short pulse of light from a GaAsP LED (1.9 eV), as described in Ref. \onlinecite{rbarends2008c}. The fibre coupling of the LED to the sample is done via a $0.15$ mm diameter waveguide that acts as a 1 THz high pass filter, to prevent any pair breaking radiation from the 4 K thermal environment where the LED is mounted to reach the sample holder \footnote{The fibreglass is opaque at 1 THz, but more transparent below 300 GHz.}. 

The complex transmission of the microwave circuit is measured with a quadrature mixer and traces out a circle in the complex plane. The microwave signal is amplified at 4 K with a high electron mobility transistor (HEMT) amplifier and with a room temperature amplifier, before it is mixed with a copy of the original signal \cite{pday2003,jbaselmans2008}. We define a resonator amplitude and phase with respect to the resonance circle, as depicted in Fig. \ref{fig:deviceandcircle}c. The resonator amplitude  predominantly responds to changes in $\sigma_1$ \cite{rbarendsphd} (therefore also called dissipation direction). The responsivity of the resonator amplitude to quasiparticles $dA/dN_{qp} = -2\alpha Q\kappa/V$, with $Q$ the quality factor of the resonator, $\alpha$ the fraction of kinetic inductance over the total inductance and $\kappa=\frac{\delta \sigma_1 / \sigma_2}{\delta n_{qp}}$, which depends only weakly on temperature. The amplitude responsivity was determined experimentally as described in Ref. \onlinecite{jbaselmans2008}. $dA/dN_{qp}$ is measured to be almost temperature independent. For similar resonators it is known that the sensitivity in phase is limited by two-level fluctuators \cite{jgao2007,jgao2008} and that the sensitivity in amplitude is up to a factor 10 better, limited by the HEMT amplifier \cite{jbaselmans2008}.

The power spectral density due to quasiparticle number fluctuations in the resonator \textit{amplitude} is given by 
\begin{equation}
	S_{A}(\omega) = S_N(\omega)\frac{\left(dA/dN_{qp}\right)^2}{1+(\omega\tau_{res})^2},
\label{eq:radiusnoise}
\end{equation}
where $\tau_{res}$ is the resonator ringtime given by $\tau_{res}=\frac{Q}{\pi f_0}$. In this experiment $\tau_r \gg \tau_{res}\approx 2$ $\mu$s, meaning that the roll-off in the noise spectrum will be determined solely by $\tau_r$ if $S_A$ is dominated by quasiparticle number fluctuations. Using Eqs. \ref{eq:Nqp}-\ref{eq:radiusnoise}, with $\tau_0=438$ ns \cite{skaplan1976} and a measured $dA/dN_{qp}=5.0\times 10^{-7}$, we expect $S_A=-99.3$ dBc/Hz. This is a high value compared to other superconductors like Ta and Nb, due to the large $\tau_0$ in Al \cite{skaplan1976}. 

\begin{figure}
\includegraphics{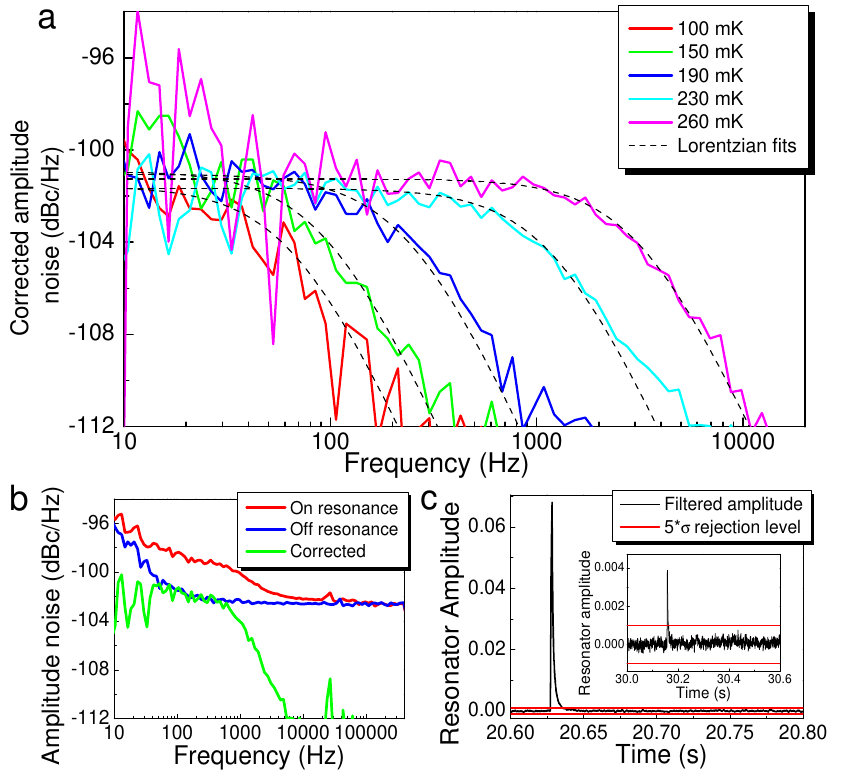} %
\caption{\label{fig:ampnoiseandcalibration} (color online) (a) Corrected power spectral density of the resonator amplitude fluctuations as a function of frequency, calibrated for system noise contributions as depicted in (b). The spectra are plotted for six different temperatures. Note that the spectral density is temperature independent up to the roll-off frequency. The Lorentzian fits to the spectra, plotted as dashed lines, show that the spectra can be described with a single timescale. (b) The noise spectrum on resonance is corrected for system noise contributions by subtracting the system noise contributions from the raw data. The system noise is obtained by taking a calibration measurement at a frequency far from the resonance frequency as indicated in Fig. \ref{fig:deviceandcircle}c. (c) Part of the time trace of the resonator amplitude. High energy impacts are observed every 20-30 s, creating millions of quasiparticles in the resonator volume. Smaller impacts, as shown in the inset, happen every 5-10 s. The time domain trace is filtered with a moving average filter with a time constant $\tau_r/2$, to distinguish small impacts from the noise. Parts of the time domain trace with impacts larger than 5 times the standard deviation are rejected, corresponding to energies of 0.43 eV or 2500 quasiparticles.} 
\end{figure}

We have measured the fluctuations in the resonator amplitude in equilibrium at the resonant frequency using a microwave power of -77 dBm. The power spectral density, corrected for system noise, is shown in Fig. \ref{fig:ampnoiseandcalibration}a for various temperatures, which is the central result of this paper. In Fig. \ref{fig:ampnoiseandcalibration}b the system noise spectrum is shown, which is subtracted from the spectrum measured on resonance, to get the corrected spectra in Fig. \ref{fig:ampnoiseandcalibration}a.
Parts of the time domain signal where large energy impacts are observed, are removed from the analysis as shown in Fig. \ref{fig:ampnoiseandcalibration}c, because they distort the dynamic equilibrium. These impacts do not appear in the system noise signal and decay within a time $\tau_r$, which shows they are due to events that create quasiparticles, for example due to cosmic ray hits \cite{lswenson2010} or local radioactivity.

In Fig. \ref{fig:ampnoiseandcalibration}a we observe that the measured power spectral density of the fluctuations has a constant level for all temperatures. The roll-off in the spectra can be described with a single timescale that decreases with temperature. From these two properties we conclude that we directly observe quasiparticle number fluctuations.

\begin{figure}
\includegraphics{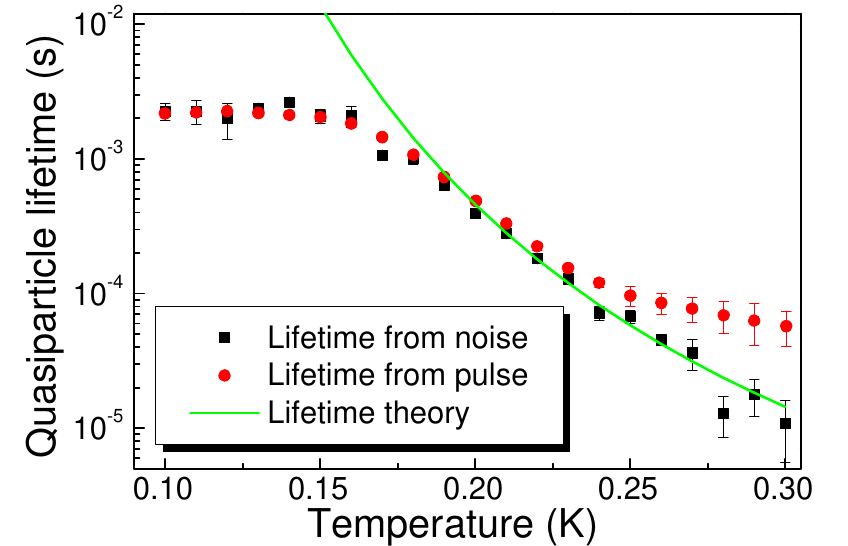} %
\caption{\label{fig:lifetimesandpulse} (color online) The quasiparticle lifetime is obtained from the roll-off frequency of the resonator amplitude noise spectrum (Fig. \ref{fig:ampnoiseandcalibration}) and plotted as a function of temperature. The solid line is the lifetime, calculated from theory. Additionally the lifetime is determined by measuring the decay of the amplitude while restoring the equilibrium after a short light pulse. The exponential decay of the excitation is fit and the obtained decay time is plotted as a function of temperature. The measurements show consistently that the lifetime saturates to about 2.2 ms below 160 mK. The error bars represent statistical uncertainties obtained from the fitting procedures.}
\end{figure}

The measured recombination time, $\tau_r$, is extracted from the resonator amplitude noise spectra and shown as the black squares in Fig. \ref{fig:lifetimesandpulse} as a function of temperature. At temperatures from 180-300 mK, we find the expected exponential temperature dependence. Eq. \ref{eq:taur} is used to fit for the characteristic electron-phonon interaction time $\tau_0$ and we find a value of 458$\pm$10 ns, in reasonable agreement with other studies \cite{skaplan1976,cchi1979}. Due to the phonon trapping effect, which we cannot estimate accurately, the measured $\tau_0$ may differ from the pure electron-phonon time \cite{skaplan1979,cchi1979}. At temperatures $<$ 150 mK we measure a temperature independent quasiparticle recombination time of 2.2 ms, which is among the longest reported for a thin superconducting film. Alternatively, the recombination time is measured by monitoring the restoration of equilibrium after a short pulse of optical photons \cite{rbarends2008c}, which is shown by the filled circles in Fig. \ref{fig:lifetimesandpulse}.  The lifetimes obtained from the noise spectra are equal to the lifetimes from the pulse measurement up to 220 mK, indicating that both measurements really probe quasiparticle recombination in equilibrium. The lifetime from the noise spectra agrees well with theory (full line) up to 300 mK. The lifetime obtained with the pulse method shows a deviation from theory, which we always observe in Al on sapphire resonators. The deviation between the two sets of experimental data calls thus, most likely, for a future analysis of the physical processes in the pulse method.

\begin{figure}
\includegraphics{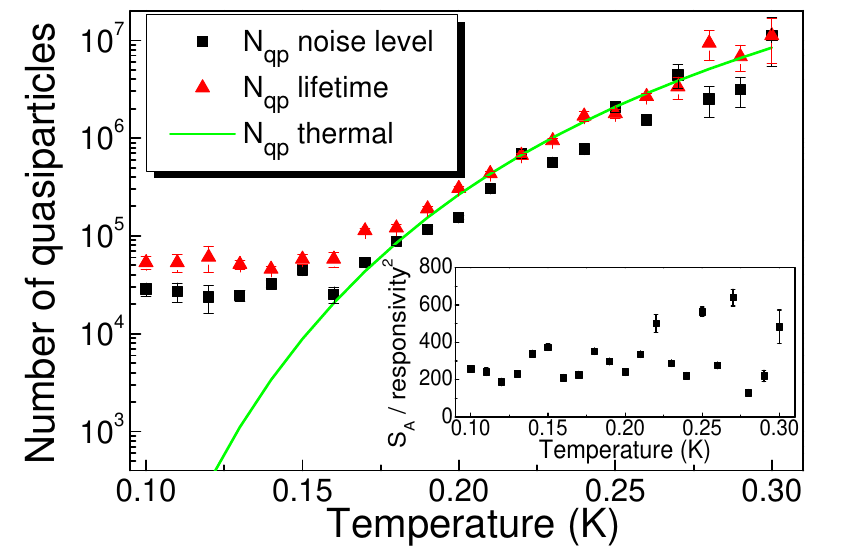} %
\caption{\label{fig:numberofquasiparticles} (color online) The number of quasiparticles as determined from the noise level as a function of temperature. The inset shows that the noise level ($S_A$) divided by the responsivity is almost temperature independent. The error bars indicate combined statistical uncertainties. Alternatively, the number of quasiparticles in the superconductor is determined from the measured quasiparticle lifetimes. We observe that the number of quasiparticles saturates at around 30,000-50,000 at low temperature, where an exponential decrease is expected, as shown by the solid line. }
\end{figure}

We combine the level of the power spectral density (Fig. \ref{fig:ampnoiseandcalibration}) \emph{and} the quasiparticle lifetime obtained from the roll-off in this spectral density (Fig. \ref{fig:lifetimesandpulse}) to obtain the number of quasiparticles, $N_{qp}$, by using Eqs. \ref{eq:SN} and \ref{eq:radiusnoise}, together with the measured values of $dA/dN_{qp}$. The result is plotted in Fig. \ref{fig:numberofquasiparticles} with the black squares. As a cross check, we convert the quasiparticle lifetime measured from the noise roll-off directly into quasiparticle number by using Eq. \ref{eq:taur}, which is shown in Fig. \ref{fig:numberofquasiparticles} with red triangles. We assume that the relevant volume is the central strip volume of the resonator. The quasiparticle number, obtained via these two methods consistently shows a saturation, giving a low temperature quasiparticle density of 25-55 $\mu$m$^{-3}$. We conclude that the quasiparticle lifetime saturation is due to a saturation in the quasiparticle density, consistent with the conjecture of Martinis et al. \cite{jmartinis2009}.

The question remains what the source of the non-thermal (in view of Eq. \ref{eq:Nqp}) quasiparticle density below 160 mK is. In literature, it is usually referred to as non-equilibrium quasiparticle density, which is inferred from saturating tunnel rates \cite{jaumentado2004,aferguson2006} and attributed to electromagnetic noise \cite{wkuo2006,bpalmer2007} and radiation \cite{onaaman2007} or, if the first two are eliminated, to cosmic rays, local radioactivity, slow heat release or stray light \cite{jmartinis2009}. In our experiment, excess quasiparticles due to cosmic ray hits are excluded as shown in Fig. \ref{fig:ampnoiseandcalibration}c. Local radioactivity is also excluded as far as it generates similar distinguishable events. The sample box and coaxial cables are thoroughly shielded from stray light, validated by the observation that $\tau_r$ is independent of the 4 K-stage temperature, as explained before. Additionally, if there would still be stray-light, it would cause an additional photon shot noise \cite{rbarends2008c} contribution, which decreases with increasing temperature, due to the decreasing quasiparticle lifetime. The observed temperature independent noise level therefore excludes stray light. We cannot completely exclude quasiparticle generation by the microwave signal. The microwave power range over which we can measure the quasiparticle fluctuations is only about 4 dB due to limitations in resonator power handling and amplifier noise temperature. Over this power range the quasiparticle density is power independent. Intriguingly, the remnant quasiparticle density is so low that a rather unexplored regime of large spatial separation between quasiparticles is entered.

For superconducting radiation detectors, the noise equivalent power (NEP) is a common way to express the sensitivity. From the noise level and lifetime measurements we determine \cite{pday2003} an electrical NEP of 3.3$\times 10^{-19}$ WHz$^{-1/2}$ due to generation-recombination noise, not taken into account the system noise, which will increase the NEP by about a factor of two. For Al resonators with this geometry and the observed remnant quasiparticle density, this is the fundamental limit to the sensitivity. A significant improvement in the NEP is only possible if one could reduce the remnant quasiparticle density or the resonator volume.

We would like to thank Y.J.Y. Lankwarden for fabricating the devices. A. Endo is financially supported by NWO (Veni grant 639.041.023) and the Netherlands Research School for Astronomy (NOVA).

\end{document}